\documentclass[reprint,twocolumn,superscriptaddress,a4paper,nofootinbib]{revtex4-2}
\usepackage{graphicx}
\usepackage{footnote}
\usepackage{bm}
\usepackage{amsmath}
\usepackage{amsfonts}
\usepackage[LGRgreek]{mathastext}
\usepackage{microtype}
\usepackage{textcomp}
\usepackage{mathtools}
\usepackage{lineno}
\usepackage{chemfig}
\usepackage{soul}
\usepackage{cancel}
\usepackage{wasysym}
\usepackage{mwe}
\usepackage{palatino}
\usepackage{textcomp}
\usepackage[section]{placeins}
\usepackage[font={small}]{caption}
\usepackage[pdfpagemode=UseNone,colorlinks=false,linkcolor=blue,citecolor=blue]{hyperref}

\begin{document}
	\title{A Novel Electrochemical Flow-Cell for Operando XAS Investigations On X-ray Opaque Supports.}
	\author{F. Paparoni}
	\email{francesco.paparoni@unicam.it}
	\affiliation{Sez. Fisica, Scuola di Scienze e Tecnologie, Universit\'{a} di Camerino, via Madonna delle Carceri, I-62032 Camerino, Italy}
	\affiliation{Synchrotron SOLEIL, L’Orme des Merisiers, BP48 Saint Aubin, 91190 Gif-sur-Yvette, France}
	\author{G. Alizon}
	\affiliation{Synchrotron SOLEIL, L’Orme des Merisiers, BP48 Saint Aubin, 91190 Gif-sur-Yvette, France.}
	\author{A. Zitolo}
	\affiliation{Synchrotron SOLEIL, L’Orme des Merisiers, BP48 Saint Aubin, 91190 Gif-sur-Yvette, France.}
	\author{A. Di Cicco}
	\affiliation{Sez. Fisica, Scuola di Scienze e Tecnologie, Universit\'{a} di Camerino, via Madonna delle Carceri, I-62032 Camerino, Italy}
	\author{S. J. Rezvani}
	\affiliation{Sez. Fisica, Scuola di Scienze e Tecnologie, Universit\'{a} di Camerino, via Madonna delle Carceri, I-62032 Camerino, Italy}
	\affiliation{CNR-IOM, SS14 – km 163.5 in Area Science Park, 34149, Trieste, Italy.}
	\author{H. Magnan}
	\affiliation{Université Paris-Saclay, CEA, CNRS, Service de Physique de l’Etat Condensé, F-91191 Gif-sur-Yvette, France}
	\author{E. Fonda}
	\email{emiliano.fonda@synchrotron-soleil.fr}
	\affiliation{Synchrotron SOLEIL, L’Orme des Merisiers, BP48 Saint Aubin, 91190 Gif-sur-Yvette, France.}
\begin{abstract}
	Improvement of electrochemical technologies is one of the most popular topic in the field of renewable energy. However, this process requires a deep understanding of the electrode–electrolyte interface behavior under \textit{operando} conditions. X-ray absorption spectroscopy (XAS) is widely employed to characterize electrode materials, providing element-selective oxidation state and local structure. Several existing cells allow studies as close as possible to realistic operating conditions, but most of them rely on the deposition of the electrodes on conductive and X-ray transparent materials, from where the radiation impinges the sample. 
	In this work, we present a new electrochemical flow-cell for \textit{operando} XAS that can be used with X-ray opaque substrates, since the signal is effectively detected from the electrode surface, as the radiation passes through a thin layer of electrolyte ($\sim$17 $\mu$m). The electrolyte can flow over the electrode, reducing bubble formation and avoiding strong reactant concentration gradients. We show that high-quality data can be obtained under \textit{operando} conditions, thanks to the high efficiency of the cell from the hard X-ray regime down to $\sim$ 4 keV. We report as a case study the \textit{operando} XAS investigation at the Fe and Ni K-edges on Ni-doped $\gamma$-Fe$_2$O$_3$ films, epitaxially grown on Pt substrates. The effect of the Ni content on the catalytic performances for the oxygen evolution reaction is discussed.
\end{abstract}
\maketitle
\section{Introduction}
\textit{Operando} investigations of the electronic and structural dynamics have a pivotal role in the development of electrochemical energy devices, such as fuel cells, electrolyzers and batteries \cite{Tikekar}. The improvement of this large family of devices requires in particular a profound knowledge of the processes taking place at the electrode-electrolyte interface in its working state. The probing of such interface under \textit{operando} conditions can be carried out by means of imaging methods, such as electrochemical scanning tunneling microscopy (EC-STM\cite{STM}) or sum frequency generation spectroscopy (SFG\cite{SFG}). However, the stringent experimental restrictions imposed by these techniques, such as low current densities and pressures\cite{SFG,wang}, complicate and limit the investigation of the catalyst performances. Another available approach is given by the vibrational spectroscopy, i.e. Raman and infrared absorption (IR). In particular, attenuated total reflection (ATR-IR) is an established technique to obtain information about electrochemical reactions at the solid-liquid interface, with the possibility to determine the species absorbed by the sample and the liquid-phase products. In some cases, even structural changes at the electrode/electrolyte can be followed by increasing the signal-to-noise ratio, exploiting the surface-enhanced infrared absorption (SEIRA) effect. However, ATR requires to deposit the sample directly on a internal reflective element, a transparent optical element with high refractive index (E.g. diamond) which can be challenging \cite{ATR, ATRdue}.
Another popular technique to probe the electronical structure of a superficial layer is X-ray photoelectron spectroscopy (XPS). In recent years, the XPS signal at near ambient pressure (NAP-XPS) have been considerably improved through the implementation of differential pumping of gasses with low cross-section, together with electrostatic focusing systems in the analyzer. Combined with a X-ray source that can provide a radiation in the tender X-ray range (2$\sim$7 keV), this new generation XPS setup allow for \textit{operando} studies of solid/gasses interfaces \cite{NAP-XPS}. The study of solid/liquid interface still requires the use of hard X-ray and an extremely thin electrolyte layer (tens of nm) \cite{AP-HAXPES}.
Recently, an increasing number of studies in the field of electrocatalysis were accomplished through the use of synchrotron-based X-ray techniques, particularly X-ray absorption spectroscopy (XAS). By combining the different penetration depth of X-ray in the soft and hard X-ray regime, both bulk and superficial region of the active material can be investigated \cite{LMO}. Soft XAS experiment under \textit{operando} conditions have been carried out using few nm thin metallic film deposited on a $\sim$ 100 nm thick silicon nitrate membrane, from where the X-ray beam enter \cite{Velasco-Velez,Schwanke,JIANG}. However, considering that this setup must be operated in vacuum conditions, the requirements for this kind of experiment can significantly complicate both sample preparation and measurement procedures. Moreover, considering that electrochemistry usually concerns phenomena occurring at the electrode-electrolyte interface, specific electrode preparations may severely deviates from the real electrodes. In the hard X-ray regime, a more straightforward approach to the operando XAS experiments on materials containing metal centers is to probe samples with high surface area to volume ratio. In this case, a single setup can provide both bulk and surface information under applied potential by analyzing both the extended X-ray absorption fine structure (EXAFS) and the X-ray absorption near edge structure (XANES) spectral region. Different design of electrochemical cells have been used, depending on the energy range and electrode type. The best option should be to probe the electrode from the face exposed to the electrolyte. However, if hard X-rays have a high penetration depth, probing lower absorption edges can be extremely challenging in the presence of aqueous electrolytes: for instance, to cross a 1 mm thick water layer a 10 keV photon flux is attenuated of the $\sim$ 41\%, while only $\sim$ 1\% of the photons can cross the liquid layer if the beam energy is 5 keV.  
In most electrochemical cells, this obstacle is usually overcome by probing the electrode from the substrate side. In this case, the sample is usually fixed on a graphite foil working electrode or glassy carbon support, since it has to be both conductive and transparent to the X-ray \cite{Volcanotrend,Zahner}. After crossing the carbon support, the beam impinges the sample in contact with the electrolyte, then the X-ray fluorescence signal is collected through the same carbon window. Another approach consists in reducing the electrolyte layer thickness. A well-known technique to this end is the so-called "dip-and-pull" method \cite{Favaro}, in which the sample is immersed in the electrolyte and then slowly extracted, achieving thin electrolyte layers of few nanometers over flat samples. However, this method is applicable only for wettable surfaces and, for faradaic reactions that involve electrolyte consumption, the mass transport limitations imposed by the extremely thin liquid layer favor the destabilization of the electrolyte\cite{dipandpull,Stabilizing}. 
Realizing an electrochemical cell where a uniform, $\mu$m thin electrolyte layer flows on the sample has proven to be quite challenging. In the system described in ref \cite{Binninger}, the authors flew a 2 mm thick liquid layer on the sample, while performing XAS measurement in fluorescence mode. This system showed excellent results at the iridium L$_3$ edge, but such thick liquid layer can be expected to severely hinder the operability at lower energies. A similar electrochemical cell setup was adopted in the work of M. Farmand et al. \cite{Farmand}. With their setup, the authors obtained a good signal at the Cu K-edge (8.98 keV), but absorption edges below 8 keV remain inaccessible. In another work \cite{XRDstudy} the authors declare to have used an extremely thin liquid layer (1 $\mu$m) in the electrochemical cell for \textit{in-situ} XRD measurements. However, even in this case, no measurements with incident energy below the Cu K-edge were presented. This suggested thin liquid layer, realized by the pressure of the electrolyte flux on a mylar window adhering to the sample, may result in a non-uniform electrolyte flow at the electrode-electrolyte interface, especially in the case of reactions involving the electrolyte consumption and formation of bubbles. 

In this work, we describe a novel electrochemical cell designed for \textit{operando} XAS measurements in fluorescence mode and compatible with acid and alkaline electrolytes. The cell can be routinely used for energies as low as Ti K edge and enable photoresponse characterization in the visible range. The low absorbance of the thin liquid layer that can be flown on top of the sample, combined with that of the window, results in a high \textit{photon-in photon-out} efficiency even in the tender X-ray range. We estimate a 43\% efficiency at the Ti K-edge and 75\% at the Fe K-edge.\\ As a case study, we discuss the \textit{operando} XAS measurements at the Fe and Ni K-edges on thin Ni-doped maghemite ($\gamma$-Fe$_2$O$_3$) films on Pt substrate. In fact, one of the hottest subjects currently investigated by \textit{operando} XAS studies is the catalysis of the oxygen evolution reaction (OER). Indeed, the sluggish kinetics of the four-electron process required to form molecular oxygen makes it essential to find high-efficiency electrocatalysts, more earth-abundant with respect to the current benchmark catalyst for OER, Ir- and Ru- based \cite{OER,OERdue}. Recently, promising results have been shown by 3d transition metal oxides like Fe$_2$O$_3$ \cite{Efficientelectrocatalyst,Constructing}. It has been suggested that the catalytic activity for OER of hematite ($\alpha$-Fe$_2$O$_3$) can be boosted by metal doping \cite{dopingofNi}. The n-type doped hematite has shown potential also for solar water splitting \cite{Fe2O3solar}. The setup described here can drastically speed up the \textit{lab-to-beamline} preparation for \textit{operando} studies, enabling the straightforward probing of samples otherwise requiring much more complex and time-consuming preparations, particularly for samples with well-defined crystal facet exposed to the electrolyte.  
\section{Experimental}	
All XAS and XRF experiments here discussed have been carried out at the SAMBA beamline (SOLEIL synchrotron). The thickness of the liquid layer was calculated with X-ray fluorescence spectroscopy. An iron (Goodfellow 99.99\%, 0.01 mm thick) and a titanium (Goodfellow 99.6\%, 0.1 mm thick) foil were mounted on the electrochemical cell. The experiment was carried out with an incident beam energy of 7.5 keV at 45° with respect to the sample surface. Ultrapure water (Milli-q) was flown into the cell at 1.7 mL/min rate. The fluorescence signal was acquired with a Canberra 35-elements monolithic planar Ge pixel array detector. Each measure consisted of five scans of 30 s acquisition time each. Spectra were averaged and corrected for the dead time. The intensities of the emission lines were then calculated by integration of the region of interest.

4 nm thick Ni-doped $\gamma$-Fe$_2$O$_3$ (111) films were epitaxially grown on Pt (001) single crystal substrates ($\diameter$ = 1 cm, thickness 1 mm) via oxygen plasma assisted molecular beam epitaxy (OPA-MBE) \cite{filmgrown}, varying the Ni content for each sample. After the deposition, samples were annealed in air at 350°C for 23 h. The Ni/Fe ratios within the samples were measured via XPS and XRF analysis (details in supporting info, S1). The results of the two techniques were averaged and reported in table \ref{samples}. \textit{As prepared} samples in table \ref{samples} were probed via XPS and XAS. Sample (*) and pristine maghemite have been probed \textit{ex-situ} and are used in the manuscript as references.
\begin{table}[h!]
	\centering
	\begin{tabular}{|cc|}
		\hline
		\textbf{Sample} &\textbf{Ni/Fe} \\
		\hline
		\textit{Ni$_{0.07}$Fe$_{2.62}$O$_4$}* & 0.028 $\pm$ 0.009 \\
		\hline
		\textit{Ni$_{0.1}$Fe$_{2.6}$O$_4$} & 0.042 $\pm$ 0.006 \\
		\hline
		\textit{Ni$_{0.53}$Fe$_{2.3}$O$_4$} & 0.23 $\pm$  0.01 \\
		\hline
	\end{tabular}   
	\caption{Stoichiometric formula of the investigated samples and relative Ni/Fe ratios obtained via XPS and XRF analyisis. Sample (*) has been used as low Ni content reference.}
	\label{samples}
\end{table}

The electrochemical characterization was performed with the cell connected to a potentiostat BioLogic SP-300. Sample were subjected to cyclic voltammetry (CV) between OCV and 0.4 V vs Ag/AgCl, until a stable response was observed. Before each measurement, the ohmic drop of the cell was measured via current interrupt method and used to apply iR compensation. The applied potentials reported in this manuscript were converted from vs. Ag/AgCl (U$_{Ag/AgCl}$) to the reversible hydrogen electrode U$_{RHE}$ with the Nernst equation (\refeq{nernst}), corrected for the uncompensated resistance (iR):
\begin{equation}
	\begin{split}
		U_{RHE} &= U_{Ag/AgCl} + U^0_{Ag/AgCl} + 0.059\cdot pH - iR \\
		&= U_{Ag/AgCl} - iR + 0.97\, V
		\label{nernst}
	\end{split}
\end{equation}
Linear sweep voltammetry (LSV) was performed at a 5 mV/s scan rate. To extract the Tafel slopes, the OER overpotential was plotted as a function of the logarithm of the absolute value of the current densities. The double layer capacitance was measured from the current values at 1.3 V vs RHE of CV at different scan rates.

\textit{Operando} XAS measurements were carried out by loading the samples on the cell's working electrode. A solution of 0.1 M KOH was flown in the cell. XAS data were collected while in chronoamperometric mode. The measure was repeated at different potentials between open circuit potential (OCV) and 2.09 V vs. RHE. After the measure at 1.12 V was carried out, the potential was reduced again to OCV and the measure repeated (OCV 2). Data were normalized to the incident photon flux and calibrated with references Fe and Ni foils. The electrolyte layer profiles above the samples were mapped by scanning along the vertical and horizontal direction with respect to the center of the hole of the Teflon mask, measuring the Ni K$\alpha$ emission line under flowing 0.1 M KOH, acquiring for 1 s a point every 0.25 mm.  

\section{Description of the Electrochemical Cell}
\begin{figure}[h!]
	\includegraphics[width=0.49\textwidth]{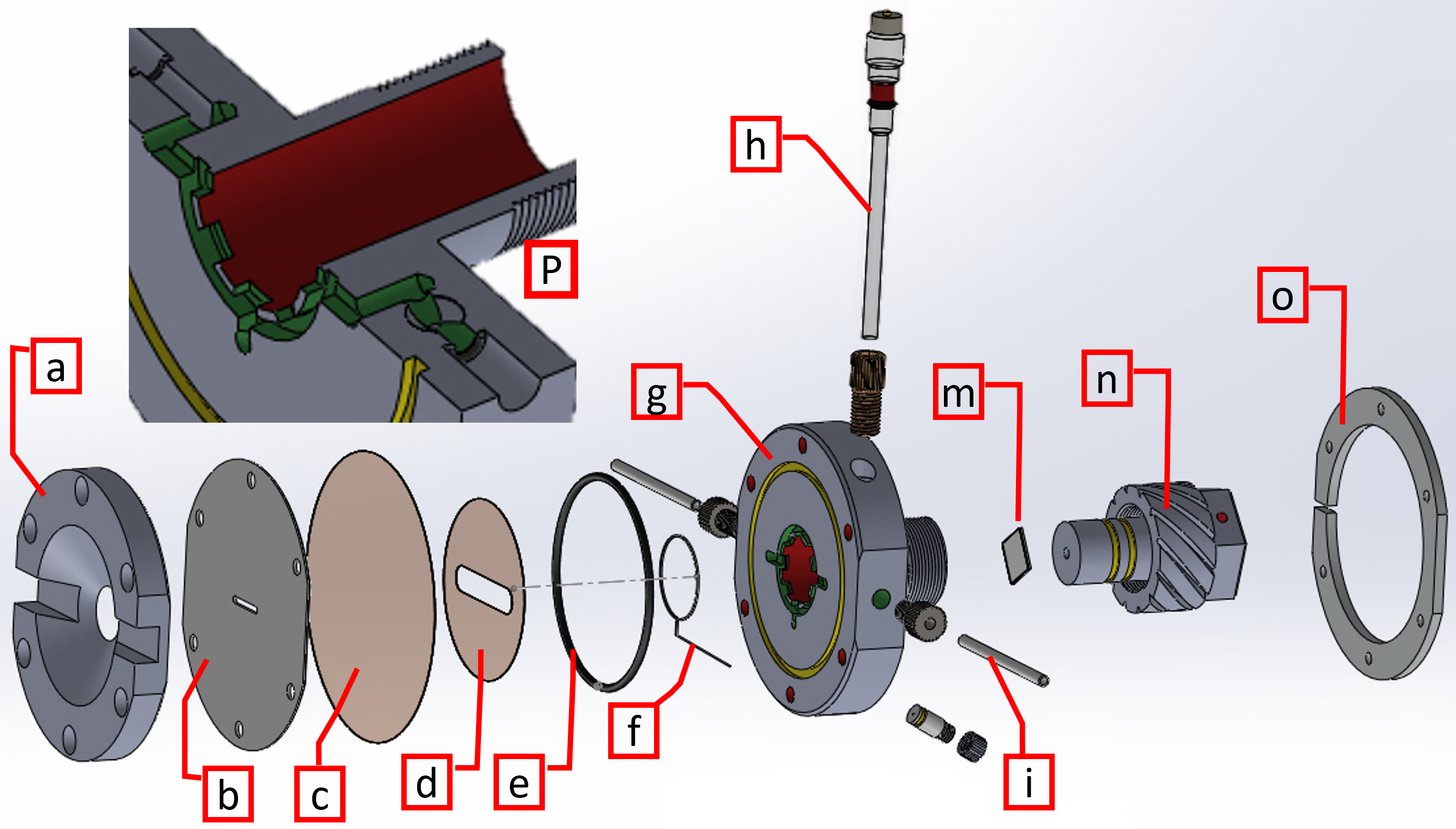}
	\caption{Exploded view of the cell: a) top PEEK plate, b) mask, c) kapton window and spacer (d), e) O-ring, f) counter electrode, g) cell body, h) reference electrode, i) electrolyte inlet (the in/out channels are carved in the counter electrode compartment, as highlighted in the inset P); l) Cell stand, m) sample n) sample holder column, o) bottom aluminum plate.}
	\label{cellexp}
\end{figure}

\begin{figure}[h!]
	\includegraphics[width=0.35\textwidth]{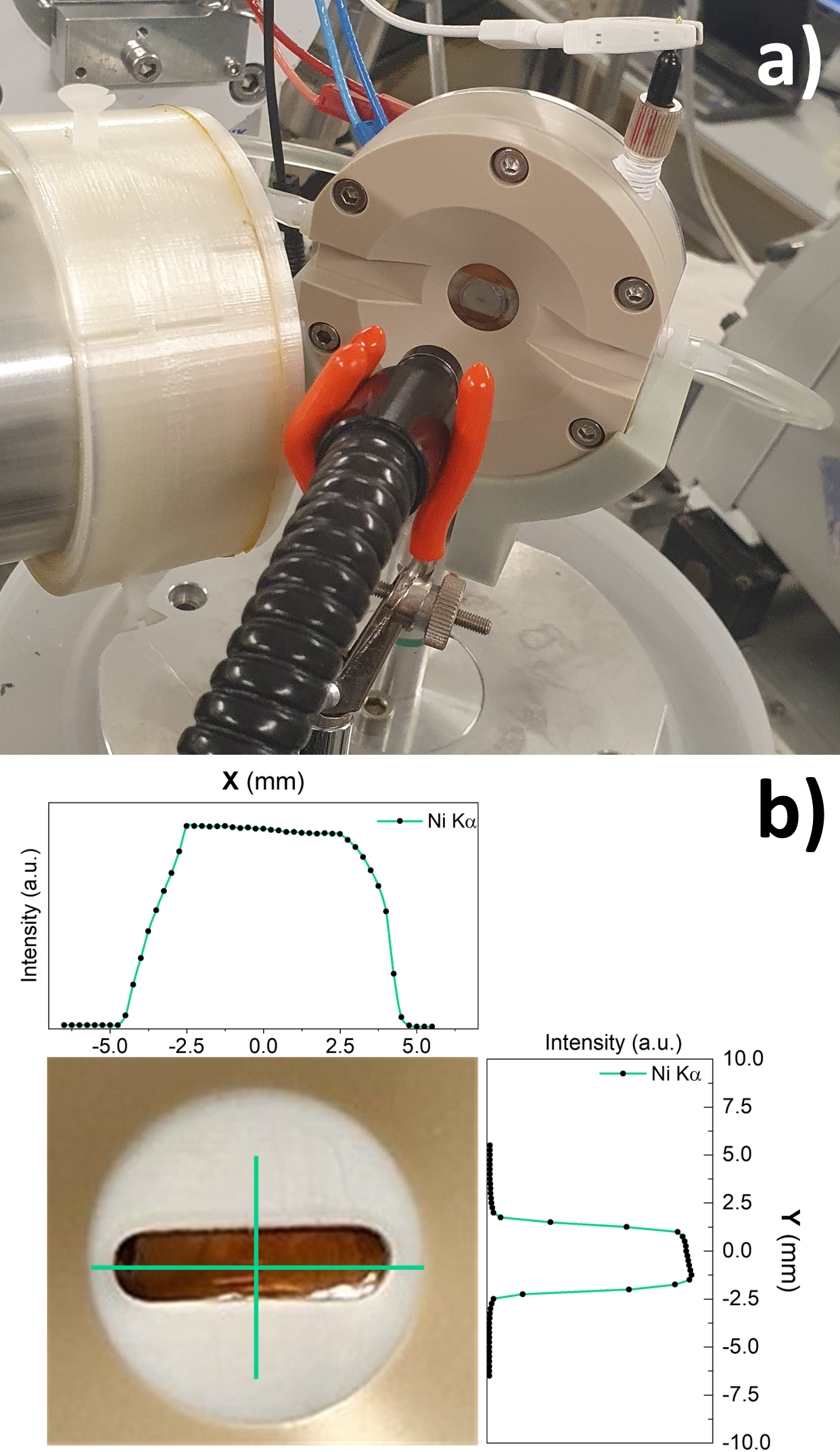}
	\caption{a) Electrochemical cell mounted at SAMBA beamline; b) inset of the cell window with the liquid layer profile curves.}
	\label{cellphoto}
\end{figure}

The cell is represented in Fig. \ref{cellexp} and consists of a circular body ($\diameter= 80$ mm) with one single circular window in the middle ($\diameter = 17$ mm).
The top plate (a) is optimized for fluorescence measurements by allowing a very large view angle of the sample (120°). Two large channels are carved to further increase this angle on the horizontal plane, where the X-ray beam incomes and where its specular diffraction may emerge (150°), letting the incidence and emerging angles to be lowered close to 15° from the sample surface. The sample is as close to the window (Fig. \ref{cellexp}c) as possible and the distance is adjustable through the sample holder (n), which consists of a retractable column that is inserted in the cell body from the back and that can be moved precisely back and forth as a screw. The sample (m) is held in place and contacted to the working electrode (Cu) from its back with PELCO\textregistered $ $ carbon conductive glue (from Ted Pella). To protect the sides of the sample and the PELCO\textregistered $ $ glue itself, nitrocellulose nail varnish is applied leaving only the front of the sample uncoated. 
The cell body (g) consists of a circular aperture in the back for the sample holding column, the aperture is surrounded by a circular compartment with a metallic Pt wire counter electrode (f) that completely surrounds the sample. The electrolyte solution enters and exits from the apertures (i), passing through channels (inset P) made to induce an electrolyte flow through before and after passing on the sample.
A reference electrode (h) placed in the flow of the electrolyte right at the beginning of the outgoing electrolyte channel completes the three electrodes cell scheme. The reference electrode chosen is a leak-free compact Ag/AgCl electrode (LF-2 model, Innovative Instruments Inc.) that guarantees alkaline and acidic media compatibility preventing electrolyte contamination. The assembled window is composed of a 25 $\mu$m thick Kapton foil and 50 $\mu$m Kapton spacer (c and d, Dupont) stuck to a 1.5 mm thick Teflon mask by wet adhesion. The mask has a 3 x 14 mm hole, aligned to the center of the spacer, from which the beam can penetrate. Top and bottom plate (a,o) press the window to the cell body (g), leaving a free volume above the sample. The electrolyte flows from inlet to outlet, filling also the counter electrode compartment (f, P) that surrounds the sample.
The body and all parts in contact with the electrolyte are made in PEEK, all o-rings sealings are made in FFKM (perfluoro elastomeric material). 
The cell is leak-tight and connected to a peristaltic pump that works pulling the liquid from the cell in normal operation, while it pushes the liquid at cell filling to remove residual gas. The pump moves the solution at a rate of 1.7 mL/min (i.e. a flow velocity above the sample of $\sim$ 0.5 m/s) renovating the electrolyte continuously, thus reducing concentration gradients and ensuring a stable cell potential. An external reservoir is placed in the circuit and it can be flushed with inert gas if needed. The thickness of the electrolyte layer above the sample can be easily controlled by rotating the sample holder column, reducing the free volume between the sample holder and the spacer. The suction produced by the peristaltic pump uniformly presses the Kapton window into the spacer and towards the sample, but keeps the liquid moving in between at the same time. This results in a uniform liquid layer above the sample (see Fig. \ref{cellphoto}) with a thickness even lower than the one of the spacer.

\section{Results}
\subsection{Thickness of the electrolyte layer}
Measuring the real thin liquid layer during cell operation is not trivial and we expected that the interplay between capillarity, hydrostatic pressure and the elasticity of the window produces a thickness above or below the spacer value, given the large (10x30 mm) clearance. Indeed, without a rigid support for the entrance window, the thickness of the liquid layer does not vary linearly with the sample distance (see supporting information, S6$^\dag$). A teflon mask (Fig.\ref*{cellexp} b), more rigid and with a smaller hole respect to the one in the spacer, helps to achieve thinner layers leaving a smaller fraction of the window self sustained.
\begin{figure}[h!]
	\centering
	\includegraphics[width=0.42\textwidth]{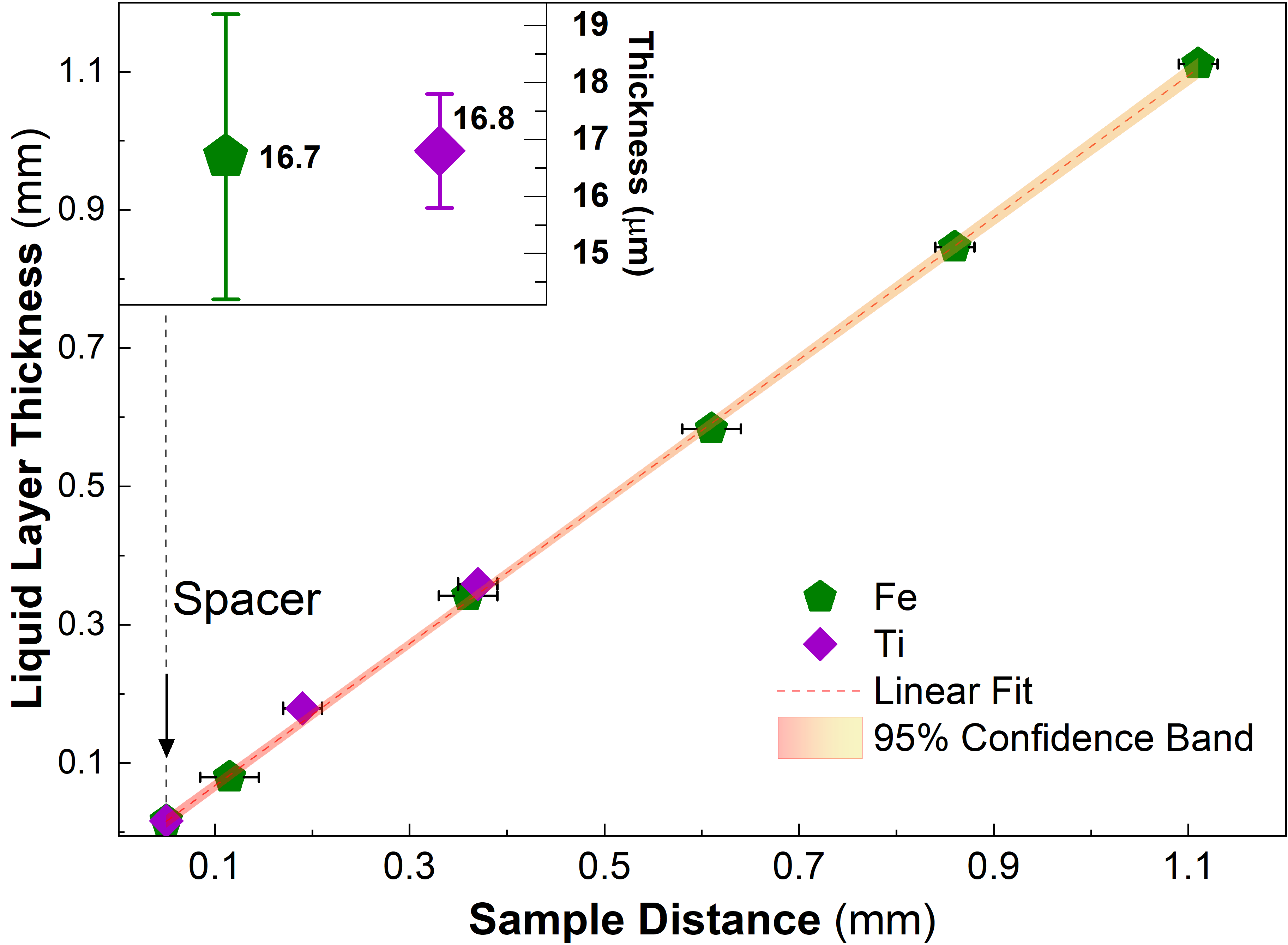}
	\caption{Liquid layer thickness vs. sample-spacer distance. The linear fit (R$^2$=0.999) was carried out on the data acquired from the Fe foil (green) and confronted with the result obtained from Ti (purple). In the inset, the minimal thickness estimated from the two measurements.}
	\label{thickness}
\end{figure}

We estimated the thickness of the liquid layer above the sample by measuring the relative intensity of the K$\alpha$ fluorescence emission of Fe and Ti foils at different distances between the sample surface and the spacer. The intensity ratio between the emission lines after (I$_1$) and before (I$_0$) the electrolyte insertion is the transmittance of the liquid layer along the optical path $L$. With $\alpha$ the angle the beam forms with the sample's surface, the liquid layer thickness ($t$) can be expressed as in eq. \ref{calcolothickness}:
\begin{equation}
	t = \frac{L}{\cos(\alpha)} = -\ln\Big({\frac{I_1}{I_0}}\Big)\cdot\frac{1}{\sqrt{2}[\mu_{in} + \mu_{out}]}
	\label{calcolothickness}
\end{equation}
with $\mu_{in}$ and $\mu_{out}$ absorption coefficients of water at the incoming (7.5 keV) and outgoing energy (4.51 keV for Ti and 6.4 keV for Fe), calculated with the \textit{XAFS mass} package\cite{Absorbix}. The flat metallic foils were fixed on the cell, at a distance of $\sim$ 1 mm from the window. After collecting I$_0$, water was flown into the cell and the measure was repeated. The initial thickness was calibrated using eq. \ref{calcolothickness}, then the measure was repeated at different sample distances. 
\begin{figure}[h!]
	\centering
	\includegraphics[width=0.42\textwidth]{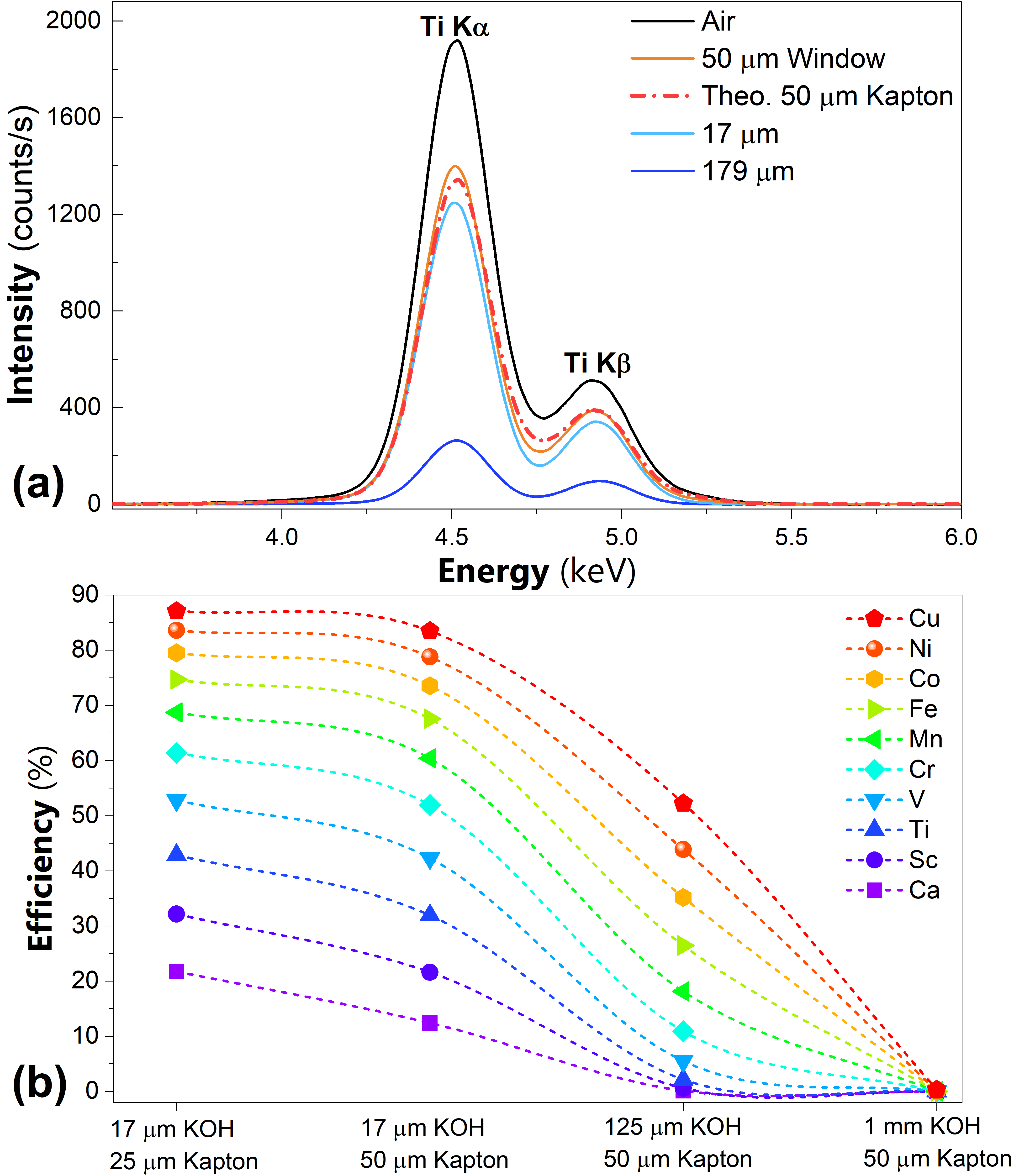}
	\caption{a) Ti K$\alpha$ and K$\beta$ acquired \textit{ex-situ} (air), with Kapton window and for two different liquid layer thickness, compared with the theoretical transmittance of 50 $\mu$m of Kapton \cite{Absorbancecalculator}; b) Theoretical efficiency of the cell for various elements, calculated for different electrolyte (KOH 0.1 M) and window thickness, assuming a probing energy 200 eV above the K absorption edge and the relative K$\alpha$ line as outgoing energy.} 
	\label{TiK}
\end{figure}
From the exponential decay of the I$_1$/I$_0$ ratios as a function of the sample distance (see supporting information, S6$^\dag$) eq. \ref{calcolothickness} confirmed a linear trend of the liquid layer thickness as the sample approaches the window, as shown in figure \ref{thickness}. 
The extremely high signal-to-noise ratio, merit of the high brilliance provided by a synchrotron light source combined with a long acquisition time, leads to a very low experimental error of the liquid layer thickness. As shown in the inset in figure \ref{thickness}, when the sample is in contact with the spacer and pressed towards the mask the results obtained from the two materials are in excellent agreement. Taking the average value of the results, we can conclude that the minimal thickness of the liquid layer achievable in this cell is 17 $\pm$ 2 $\mu$m.

The total \textit{photon-in photon-out} efficiency of the cell can be evaluated as the product of the transmittance of the liquid layer with the one of the window. To this end, the K$\alpha$ emission line of the Ti foil was measured with and without a 50 $\mu$m Kapton window. As shown in figure \ref{TiK}(top), the signal drop caused by the Kapton window (44\%) is in excellent agreement with the theoretical value expected for a 50 $\mu$m Kapton window in a 45° configuration, with incoming energy of 7.5 keV and outgoing energy of 4.51 keV. The total efficiency of the cell can be further improved by mounting a thinner Kapton window (e.g. 25 $\mu$m), as shown in figure \ref{TiK}(bottom). Our results highlight the net increase in efficiency achieved even at moderately high energies by reducing the electrolyte layer thickness.
\subsection{Electrochemical Measurements}
The reliability of the cell for electrochemical studies was tested by comparing CV measurements with published results, as detailed in supporting information (S3$^\dag$). Nonetheless, in a system with variable geometry and thin electrolyte layer such as the one here described, Ohmic overpotential can become a relevant source of error, inducing an overestimation of the applied potential due to the uncompensated resistance between the working and the reference electrode. This resistance is mostly due to the cell geometry and solution conductance \cite{ohmicdue} and, when necessary, it can be accounted for with a careful iR compensation (see 4S$^\dag$). All following results are iR-compensated.

\begin{figure}[h!]
	\centering
	\includegraphics[width=0.42\textwidth]{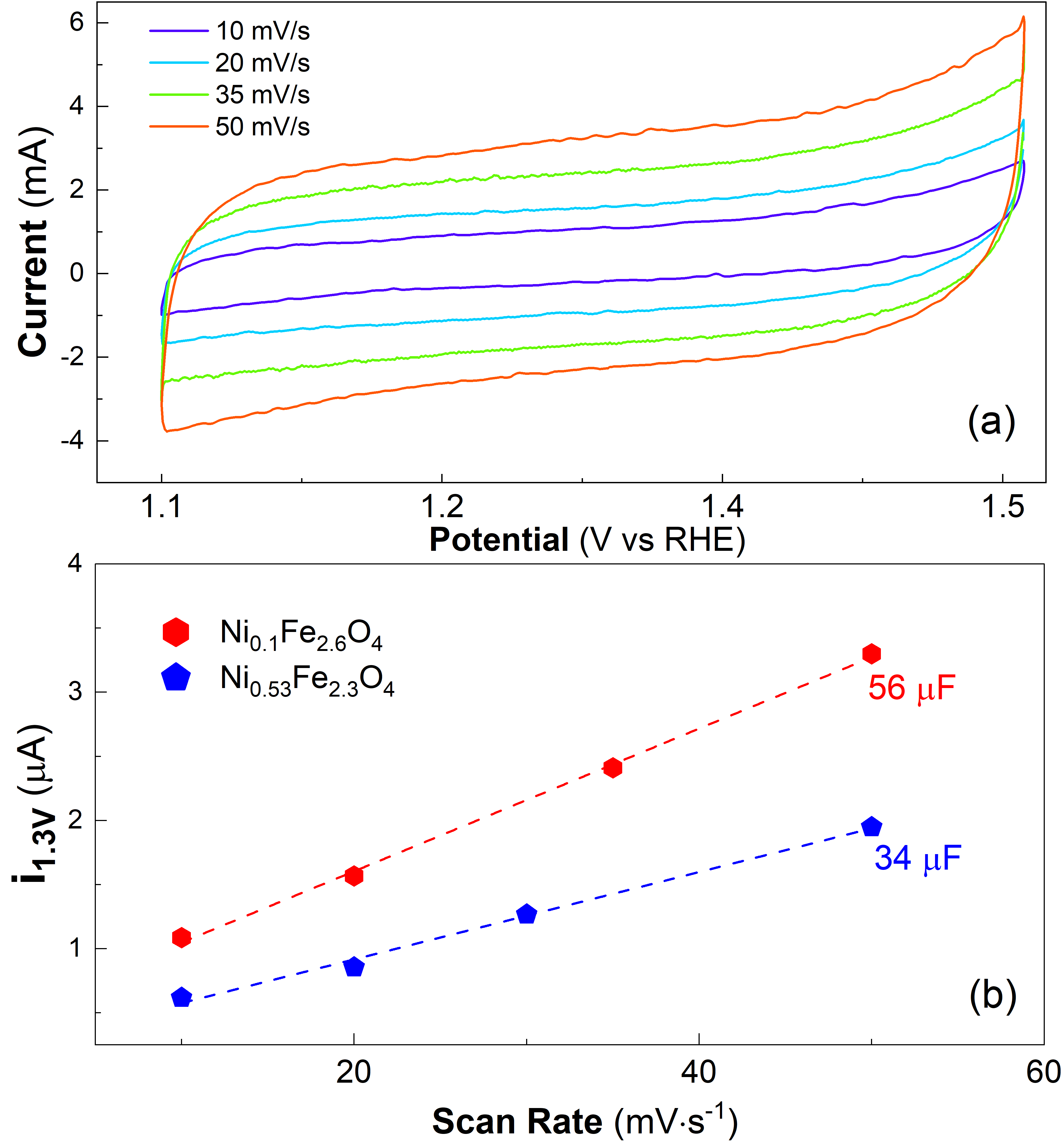}
	\caption{a) CV at different scan rates on Ni$_{0.1}$Fe$_{2.6}$O$_4$ and relative linear fit of the current at 1.3 V vs RHE (b). The obtained values of double layer capacitance are reported on the graph with corresponding colors.}
	\label{CV}
\end{figure}
\begin{figure}[h!]
	\centering
	\includegraphics[width=0.42\textwidth]{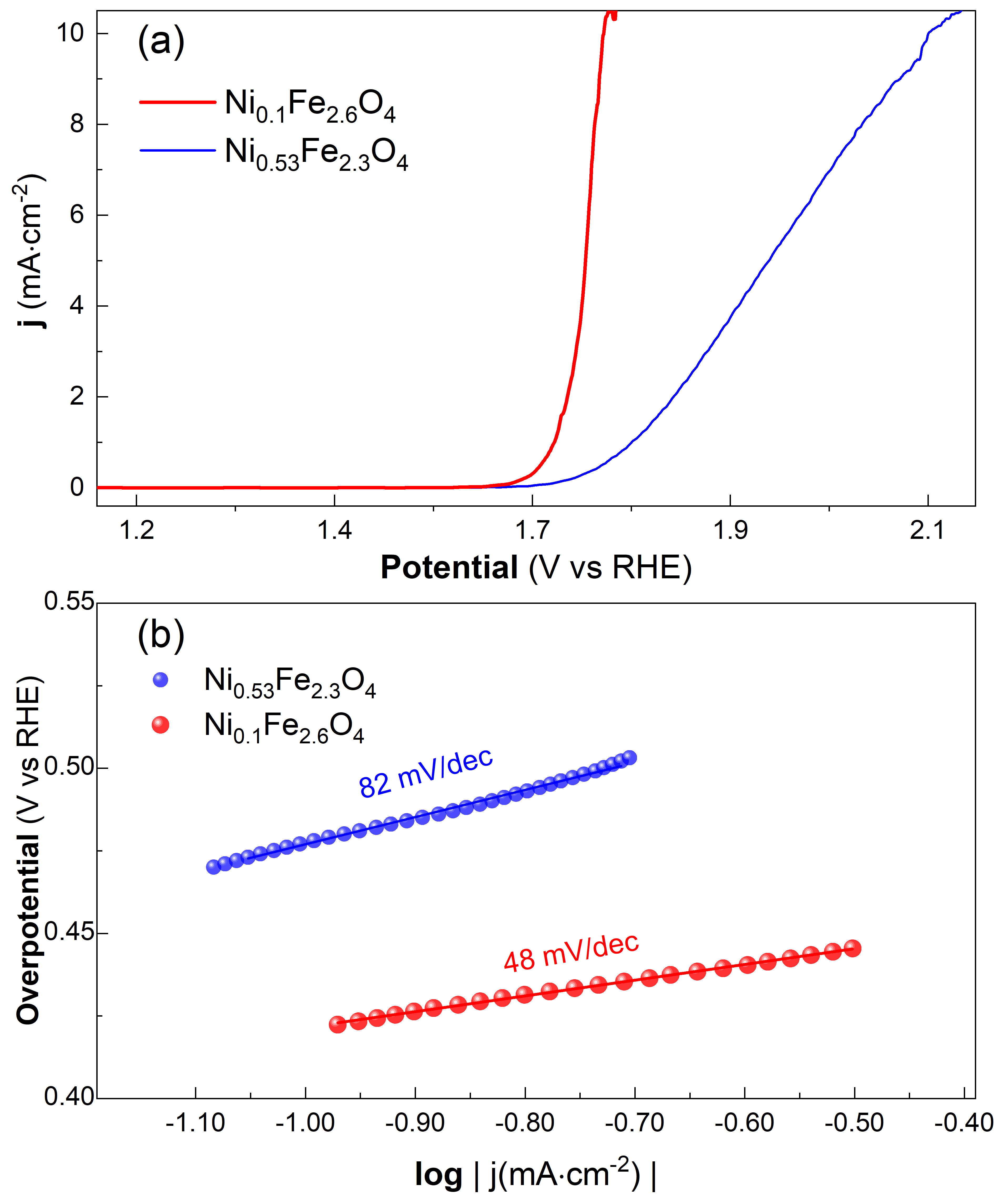}
	\caption{iR corrected LSV (a) with relative Tafel plot slopes (b). The noise above 10 mA$\cdot$cm$^{-2}$ is due to the formation of a significant amount of oxygen bubbles.}
	\label{tafel}
\end{figure}

The role of Ni-dopant concentration in the catalytic performances of maghemite for the OER was investigated via CV and LSV. As shown in figure \ref{CV}a, samples were cycled at different scan rates in a non-faradaic potential region. Following the protocols discussed in other works\cite{ohmicdue,CDL} the double layer capacitance (C$_{DL}$) was extrapolated from the linear fit of the current as a function of the scan rate ($v$). The electrochemical active surface area (ECSA) can be calculated from the following equation:
\begin{equation}
	ECSA = \frac{i}{v\cdot C_s} = \frac{C_{DL}}{C_s}
	\label{ECSA}
\end{equation}
with $C_s$ specific capacitance, that for our sample in an alkaline environment can be estimated to be 40 $\mu$F$\cdot$ cm$^{-2}$. Equation \ref{ECSA} yields a ECSA of 1.4 cm$^2$ for Ni$_{0.1}$Fe$_{2.6}$O$_4$ and 0.85 cm$^2$ Ni$_{0.53}$Fe$_{2.3}$O$_4$. Figure \ref{tafel}a shows the polarization curves obtained from a sample with low (red) and high (blue) Ni content.
The sample with 4\% Ni/Fe ratio shows the best OER performance, with a lower overpotential to reach 10 mA$\cdot$cm$^{-2}$ (0.54 V for Ni$_{0.1}$Fe$_{2.6}$O$_4$, 0.89 V for the Ni-rich sample) as well as Tafel slope (Fig. \ref{tafel}b), suggesting a more facile kinetics of oxygen exchange for Ni$_{0.1}$Fe$_{2.6}$O$_4$.

\subsection{XAS Measurements}
\begin{figure}[h!]
	\centering
	\includegraphics[width=0.42\textwidth]{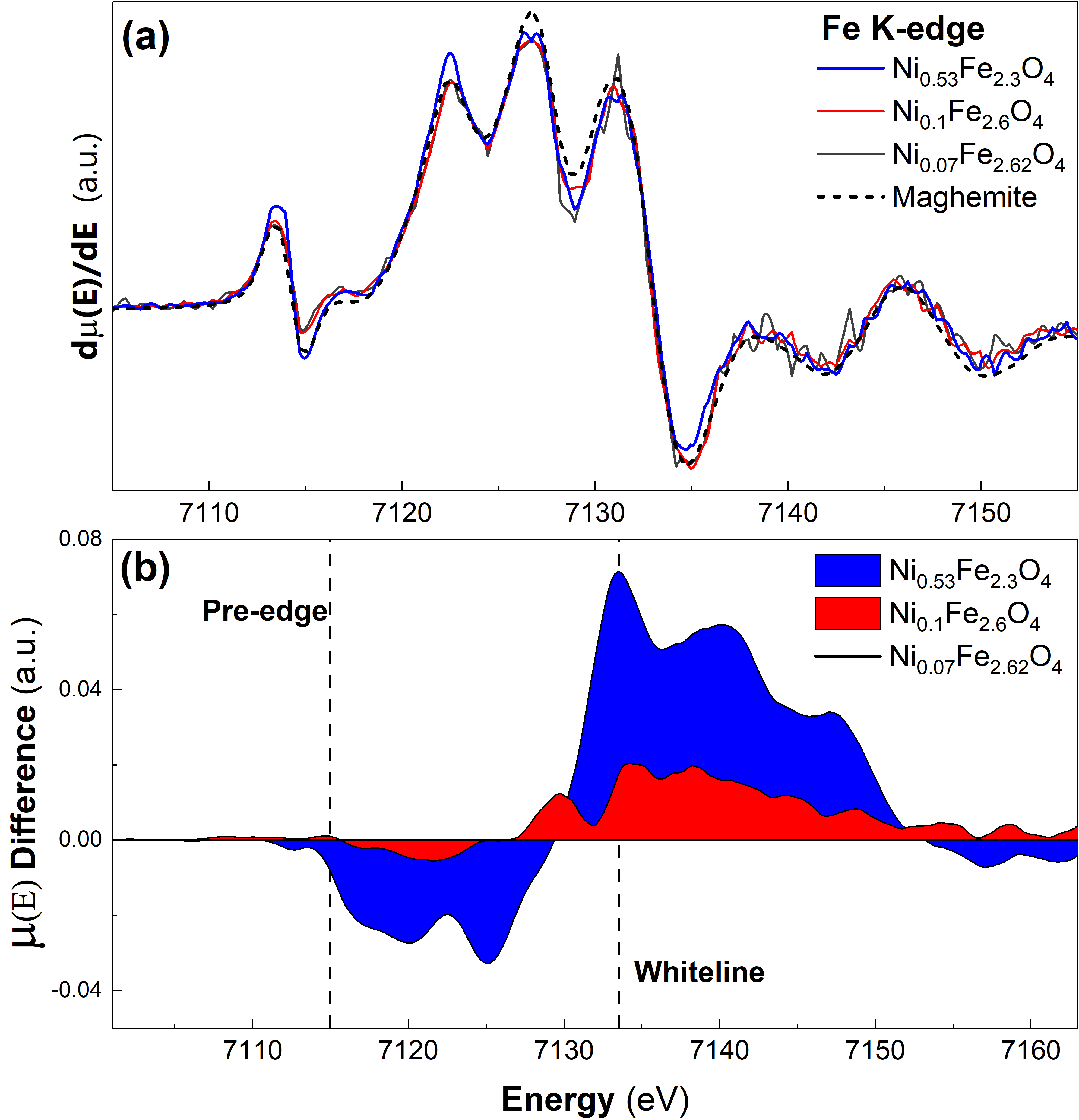}
	\caption{a) First derivative of the Fe K edge XANES at OCV Vs pristine maghemite; b) Spectral difference of the XANES data with respect to the Ni$_{0.07}$Fe$_{2.62}$O$_4$ reference.}
	\label{Fe-XANES}
\end{figure}

X-ray absorption spectra at the Fe and Ni K-edges were acquired at different operating potentials between OCV and 2.09 V vs RHE. At OCV, samples show very similar Fe K-edge spectra with respect to the maghemite reference (see Fig. \ref{Fe-XANES}). In agreement with the work of M. Coduri et al. \cite{Localstructure}, the first derivative of the Fe K-edge spectra highlights the presence of a pre-edge peak as well as three peaks at 7122, 7127 and 7130 eV. These features confirm the absence of hematite side-products, that could form in the case of the substitution of Fe ions in tetrahedral sites (Fe$_{th}$) by Ni ions. The preservation of the cubic structure in the case of the site-specific doping (i.e. Fe$^{3+}$ in octahedral sites) has been already observed by J. Chen et al. \cite{Tidoping}. 
As shown by the spectral difference in figure \ref{Fe-XANES}b, samples with more Ni content show higher intensity at the white line and a hindering in the region between the peak at 7130 eV and the pre-edge, resulting in a sharper peak around 7114 eV.
\begin{figure}[h!]
	\centering
	\includegraphics[width=0.42\textwidth]{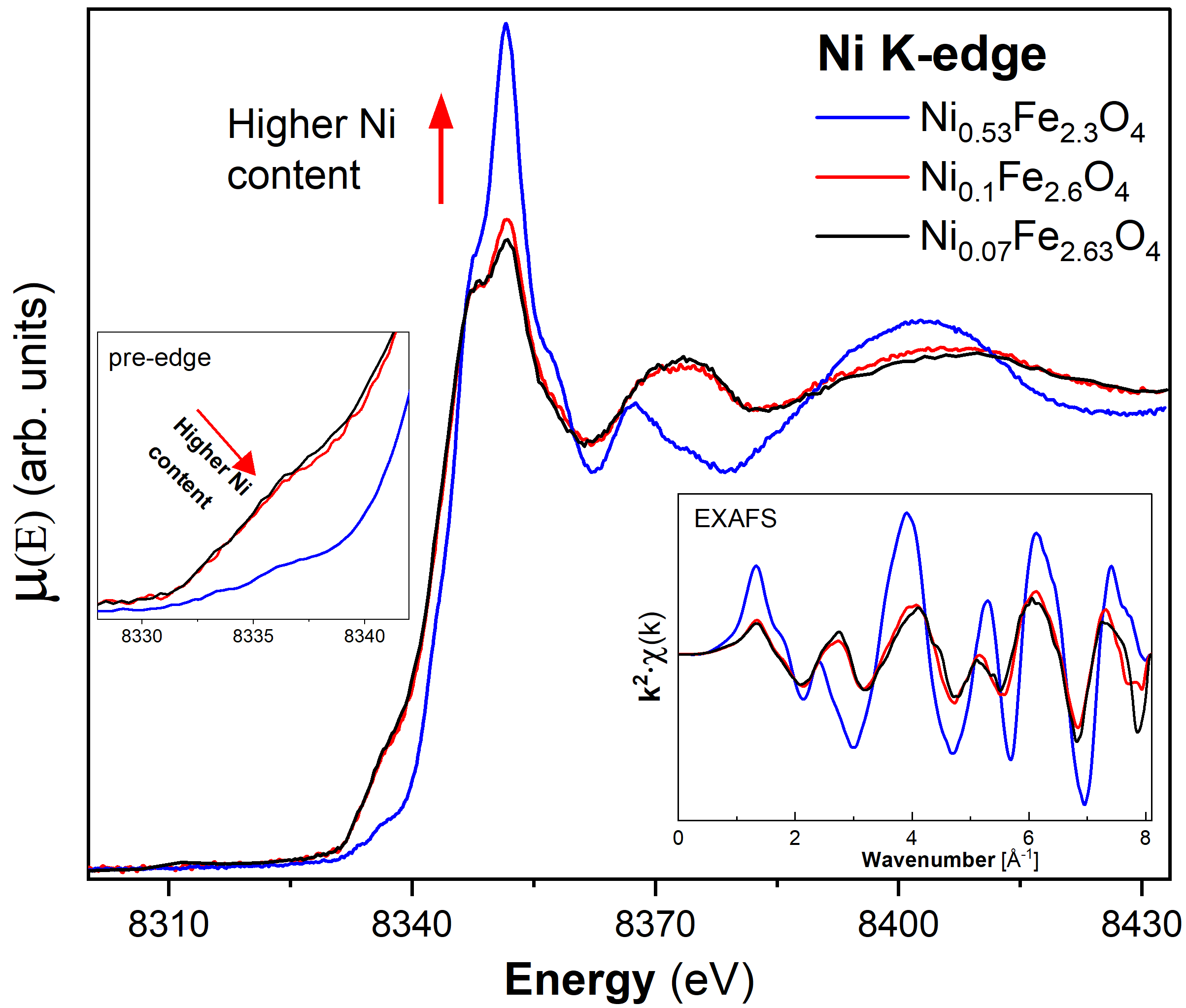}
	\caption{XANES of the Ni K-edge at OCV. In the insets, the pre-edge regions (left) and the EXAFS signal.}
	\label{Ni-XANES}
\end{figure}
Maghemite is known to exhibit a sharper pre-edge compared to other Fe$_2$O$_3$ phases like Hematite, due to the presence of tetrahedrally-coordinated Fe$^{3+}$ ions \cite{de_Groot,fitpreedge}. 
This peak may be assigned both to quadrupolar (1s $\rightarrow$ 3d) and/or more intense dipolar transitions (1s $\rightarrow$ 3d) permitted through hybridization of Fe 3d with O 2p levels \cite{Henderson,hybridization} in non-centrosymmetric geometries.
The intensity and position of this feature give information on the structural order and oxidation state of the photoabsorber. Following the procedure described by Boubnov et al. \cite{fitpreedge}, the pre-edge intensity and position were fitted with a single pseudo-voigt, subtracting the edge onset with an arctangent function (see supporting info, S4$^\dag$). The gaussian fraction (99\%) and full width at half maximum (2.6 eV) were fixed to the values reported for maghemite. The position obtained (7114.3 $\pm$ 0.1 eV) confirms the expected oxidation state of iron (Fe$^{3+}$), in agreement with XPS measurements (see supporting information, S1$^\dag$). The pre-edge intensity increases as a function of the Ni content, resulting in a sharper feature in sample Ni$_{0.53}$Fe$_{2.3}$O$_4$.\\
At the Ni K-edge, higher Ni content results in an evident increase of the white line intensity, a flattening of the pre-edge region as well as a clear modification of the fine structure (see Fig. \ref{Ni-XANES}).
As detailed in supporting information (S4$^\dag$), using as references the Ni-poor sample (Ni$_{0.07}$Fe$_{2.62}$O$_4$) and the pristine NiFe$_2$O$_4$ spectra reported by Q. Yue et al. \cite{NiFe2O4-vac}, we observe a much higher contribution (87\%) of the former compound in Ni$_{0.53}$Fe$_{2.3}$O$_4$. This result suggests a gradual ordering of the structure towards NiFe$_2$O$_4$ as the Ni concentration increases.

\begin{center}
	\begin{figure}[b]
		\includegraphics[width=0.97\textwidth]{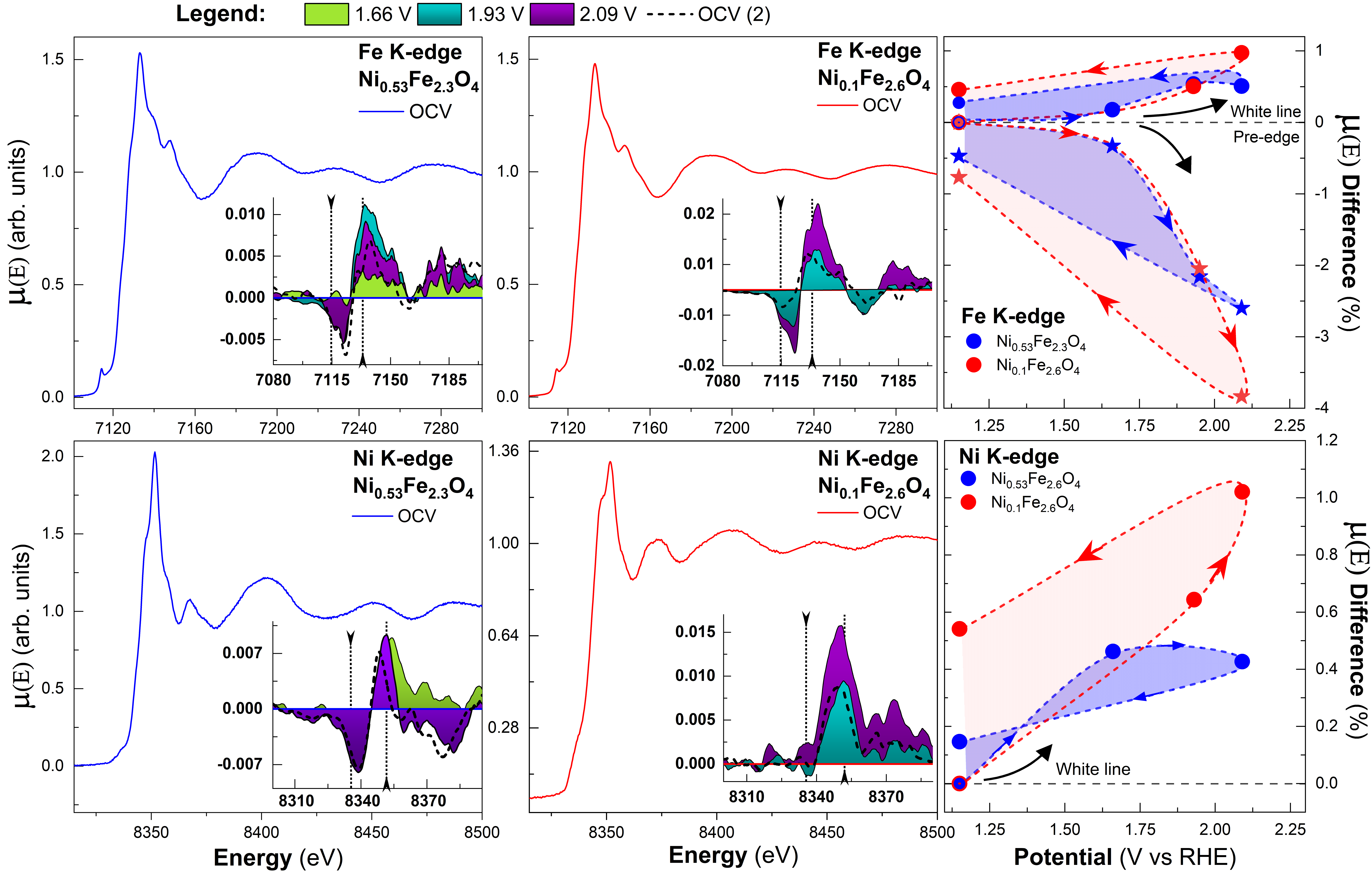}
		\caption{XANES at the iron (first row) and nickel (second row) K-edges at open circuit potential in 0.1 M KOH solution. In the insets, the relative $\mu$(E) differences between the spectra acquired at each potential point and the shown spectra (OCV) are plotted as colored areas. The dashed lines named OCV (2) indicate the spectral difference between the OCV spectra after the cycle and the initial one. Vertical dotted lines highlight the pre-edge and white line positions. On the right the results of eq. \ref{eqint} show the trend occurring at the Fe and Ni white lines (dot) and Fe pre-edge (star). Arrows indicate the chronological order of acquisition.}
		\label{XANES}
	\end{figure}
\end{center}

While changing the applied potential did not clearly affect the EXAFS signal, a potential-dependent modulation of the white line and pre-edge intensities at both edges was observed above 1.66 V. Due to the relatively low variations, these modifications of the XANES were characterized by studying the integral of the spectral difference, in a similar fashion to the protocol described in ref \cite{intmethod}. With $I_V$ absorption coefficient acquired at the potential V and $I_{OCV}$ the one acquired at open circuit potential, the percentage variation of the white line intensity ($\Delta$$WL$) is the integral around the white line energy (E$_w$):
\begin{equation}
	\Delta WL (\%)	= 100\cdot\int_{E_w-4}^{E_w+4} \frac{I_V(E)-I_{OCV}(E)}{I_{OCV}(E)} dE
	\label{eqint}
\end{equation}
The Fe K pre-edge modulations where calculated from the evolution of the peak heights, fitted with the same method previously described. As shown in figure \ref{XANES}, at the Fe K-edge both samples show a gradual increment of the white line intensity above 1.66 V. 
\onecolumngrid
\twocolumngrid
An opposite trend is observed at the pre-edge. The relative changes are higher for the sample with lower Ni content at the maximum potential (2.09 V).
When the potential is lowered back to OCV, the decrease of the white line and re-increase of the pre-edge intensities to values closer to the initial ones suggest a partial reversibility of the process. The same white line trend is observed at the Ni K-edge, with a maximum increase of $\sim$ 1\% on Ni$_{0.1}$Fe$_{2.6}$O$_4$ and $\sim$ 0.5\% on Ni$_{0.53}$Fe$_{2.3}$O$_4$ and partial reversibility after the first cycle. In this case the pre-edge modulations were not quantified, since the very broad pre-edge features of Ni$_{0.1}$Fe$_{2.6}$O$_4$ thwart an accurate quantification.

\section{Discussion}	
The structure of pure Maghemite can be described as Magnetite (Fe$_3$O$_4$) with octahedral iron vacancies. Maghemite has an inverse spinel structure with cubic symmetry (P4$_3$2) with eight Fe$^{3+}$ cations occupying tetrahedral sites (Fe$_{th}$) and sixteen octahedral sites (Fe$_{oh}$) filled by Fe$^{3+}$ except for 1/6 of vacancies, resulting in a stoichiometric formula of Fe$^{3+}$$_{(th)}$Fe$^{3+}$$_{1.66(oh)}$V(Fe)$_{0.33(oh)}$O$_4$, with V(Fe) octahedral iron vacancies, or Fe$_{2.667}$O$_4$ \cite{magnetite-struct}.
It was observed that doping of maghemite with transition metals like Co and Ti results in the enhancement of the cubic phase structural order by replacing Fe$^{3+}$ and/or Fe$_{oh}$ vacancies at the octahedral sites \cite{Magneto-thermal,Tidoping}. Hence, considering the similar ionic radii and electronegativity of octahedrally-coordinated Ni$^{2+}$ (0.69 $\AA$, 1.91) and Fe$^{3+}$ (0.64 $\AA$, 1.9), low content of Ni atoms can be expected to fill the Fe$_{oh}$ sites, improving the conductivity of maghemite without distorting the cubic structure \cite{Localstructure}. Our Fe K-edge data at OCV agrees with the preservation of the maghemite structure. The increase of the white line intensities and the sharpening of the pre-edges at the Fe and Ni K-edges as a function of the Ni/Fe ratio suggest the gradual filling of the octahedral vacancies by Ni$^{2+}$ ions, resulting in the higher structural order displayed by the Ni-rich sample (Ni$_{0.53}$Fe$_{2.3}$O$_4$). Considering also the increasing correlation between the XANES at the Ni K-edge and a NiFe$_2$O$_4$ reference, the higher Fe-O bond distances and Debye-Waller factors for lower Ni content (supporting info, S5$^\dag$) our results suggest that the substitution of Ni into the octahedral sites results in the transition towards NiFe$_2$O$_4$, in agreement with the work of Q. Yue et al.\cite{NiFe2O4-vac}. Our analysis does not evidence any change in oxidation state for Ni or Fe. Hence, the incorporation of Ni results in an intermediate phase with the stoichiometric form Fe$^{3+}$$_{(th)}$Fe$^{3+}$$_{1+2x(oh)}$Ni$^{2+}_{1-3x(oh)}$V(Fe)$_{x(oh)}$O$_4$ that retains the inverse spinel structure with cubic symmetry. While the XAS data of the sample Ni$_{0.53}$Fe$_{2.3}$O$_4$ are in excellent agreement with their NiFe$_2$O$_4$ reference, samples with lower Ni/Fe ratios show a much more defective structure at the Ni K-edge. The less evident differences observed at the Fe K edge can be attributed to the higher probability of the formation of a vacant site at the Ni neighbors, due to the weakness of the Ni-O bond with respect to the Fe-O \cite{NiFe2O4-vac}.\\
When the applied potential is increased, both samples highlight a gradual increase of the white line intensity at both edges, together with a lowering of the pre-edge intensity at the Fe K-edge. The raising of the white line evidences an increased number of unoccupied d-states for both Fe and Ni ions as a function of the applied potentials \cite{whiteline}. In the previously mentioned work \cite{NiFe2O4-vac}, the authors correlated the increase of the white line and decrease of the pre-edge with a diminution of the concentration of oxygen vacancies in NiFe$_2$O$_4$. Therefore, we can conclude that applied potential reduces the oxygen vacancy concentration, either by direct oxidation or by absorption of species. These phenomena can be understood considering the role of oxygen vacancy in the catalytic performances of NiFe$_2$O$_4$ for OER. Q. Yue et al. showed that OER is favored for higher concentrations of oxygen vacancies \cite{NiFe2O4-vac}. This correlation is not only due to the increase of the conductivity of the material, but also to the increasing of the number of active sites. Via Density-functional theory, they demonstrated that the absorption energy of H$_2$O molecules on the vacant sites is lower than the one on the stoichiometric surface and that the absorbed water is thermodynamically favored to split into OH and H releasing one electron. It is known that the OER in alkaline media will start with the absorption of a OH$^-$ from the electrolyte to active site \cite{OERsteps,OERalkaline} (eq. 5). In the adsorbate evolution mechanism (AEM), the second step of the OER (eq. 6) consist in the further oxidation of the hydroxyl at the active site (*) that will then absorb another hydroxyl from the electrolyte (eq. 7), before releasing a oxygen molecule as a final step (eq. 8):
\begin{eqnarray}
	4OH^- \rightarrow OH^* + 3OH^- + e^- \\
	OH^* + 3OH^- \rightarrow O^* + 2OH^- + H_2O + e^- \\
	O^* + 2OH^- + H_2O \rightarrow OOH^* + OH^- + H_2O + e^- \\
	OOH^* + OH^- + H_2O \rightarrow O_2 + 2H_2O + e^-
	\label{OER}
\end{eqnarray}
The insertion of hydroxyl anions in the superficial vacant sites cause the diminution of O vacancy content and upshift of the d-band center of Fe and Ni neighbors, leading to the depopulation of antibonding states, resulting in the increase of the white line at both edges and decreasing of the Fe pre-edge intensity \cite{dband}. The higher pre-edge and lower white line shown by the sample with 4\% Ni suggest a higher O vacancy content with respect to the sample with 23\% of Ni. Hence, a superior number of active sites can be expected on sample Ni$_{0.1}$Fe$_{2.6}$O$_4$, with a consequently higher amount of hydroxyl absorbable. This is in agreement with the higher variations displayed by this sample as a function of the applied potential (see figure \ref{XANES}). Our results suggest that the potential modulates O vacancies number through hydroxyl absorption.

Our electrochemical analysis confirms the superior catalytic activity of the sample with lower Ni content. Tafel analysis suggests that sample Ni$_{0.1}$Fe$_{2.6}$O$_4$ presents a lower overpotential to reach a current density of 10 mA$\cdot$cm$^{-2}$ (540 mV for against the 890 mV for Ni$_{0.53}$Fe$_{2.3}$O$_4$) as well as Tafel slope (48 mV/dec instead of 82 mV/dec). Moreover, by measuring the sample's C$_{DL}$ we estimated a larger ECSA (1.4 cm$^2$) for sample Ni$_{0.1}$Fe$_{2.6}$O$_4$. These results are in agreement with the same work already mentioned \cite{NiFe2O4-vac}, with the Tafel slope of sample Ni$_{0.53}$Fe$_{2.3}$O$_4$ in excellent agreement with the one reported for pristine NiFe$_2$O$_4$, while the slope of Ni$_{0.1}$Fe$_{2.6}$O$_4$ is close to the one reported for a NiFe$_2$O$_4$ with high concentration of oxygen vacancies. The lower overpotentials reported in the same reference could be ascribed to the mesoporous nature of their samples, while our epitaxially grown samples expose only a well-defined facet (111).  

\section{Conclusions}  
In this work, we described a novel electrochemical cell for \textit{operando} XAS investigations on X-ray opaque supports. We showed via XRF spectroscopy that we can control the volume of electrolyte flowing over the sample, down to a minimal thickness of $\sim$ 17 $\mu$m. The cell shows excellent efficiency for fluorescence experiments even in the tender X-ray range, with a total transmittance of 43\% at the Ti K-edge. As a case study, we investigated Ni-doped $\gamma$-Fe$_2$O$_3$ as catalyst for the OER as a function of the Ni content. Our \textit{operando} XAS investigations highlighted a superior catalytic activity for the sample with lower Ni content. The high quality of the XAS data allowed us to study small modulations of the XANES, attributed to the absorption and oxidation of hydroxyl species at the active sites. These results shed light on the interesting properties of Ni-doped maghemite. Further studies are required to optimize the doping content and investigate the role of surface orientation.\\ 
The high efficiency of this cell, combined with the brightness and high energy resolution provided by a synchrotron facility, allows for \textit{operando} XAS measurements at lower energies with respect to what has been reported so far (to the best of our knowledge) for other setup working with the same geometry. By expanding the list of accessible absorption edges, in a geometry that has little to no restrictions in terms of substrate requirements, this novel electrochemical cell provides the opportunity to study new functional materials via \textit{operando} X-ray absorption spectroscopy. 

\section*{Author Contributions}
F.P. performed all measurements, calculations and data curation. E. F., G.A. and A.Z. conceptualized and designed the cell; E.F. and A.Z. supervised XAS measurements; E.F. supervised the calculations and data curation. H.M. prepared the samples and participated in XAS measurements. All authors contributed to the final editing of the manuscript.
\section*{$\dag$~Supplementary Information (SI)}: SI available. Ni content calculations (S1); Operando XRF (S2); Electrochemical Tests (S3); XANES (S4); EXAFS (S5); Electrolyte thickness (S6).
\section*{Conflicts of interest}
The authors have no conflict of interest to declare.
\section{Acknowledgments}
The authors thanks the ANR for financial support through Project ANR-20-ce42-015.
\bibliography{Cell_bib}
\end{document}